\documentclass[usenatbib]{mnras}

\usepackage{amsmath}
\usepackage[final]{graphicx}
\usepackage{amssymb}
\usepackage{color}
\usepackage{hyperref}

\renewcommand{\vec}{\mathbf}

\title[Gas production from comet 67P/C-G]{The gas production of 14 species from comet 67P/Churyumov-Gerasimenko based on DFMS/COPS data from 2014-2016}
\author[Matthias L\"auter et al]{
  Matthias L\"auter,$^{1}$\thanks{E-mail: laeuter@zib.de} Tobias
  Kramer,$^{1,2}$ Martin Rubin,$^{3}$ Kathrin Altwegg$^{3}$ \\
  $^{1}$Zuse Institute Berlin, 14195 Berlin, Germany\\
  $^{2}$Department of Physics, Harvard University, Cambridge, MA, USA\\
  $^{3}$University of Bern, Physikalisches Institut, 3012 Bern, Switzerland}

\date{Accepted XXX. Received YYY; in original form ZZZ}

\pubyear{2020}

\begin{document}
\label{firstpage}
\pagerange{\pageref{firstpage}--\pageref{lastpage}}
\maketitle

\begin{abstract}
The coma of comet 67P/Churyumov-Gerasimenko has been probed by the {\it Rosetta} spacecraft and shows a variety of different molecules.
The ROSINA COmet Pressure Sensor and the Double Focusing Mass Spectrometer provide {\it in-situ} densities for many volatile compounds including the 14 gas species H$_2$O, CO$_2$, CO, H$_2$S, O$_2$, C$_2$H$_6$, CH$_3$OH, H$_2$CO, CH$_4$, NH$_3$, HCN, C$_2$H$_5$OH, OCS, and CS$_2$.
We fit the observed densities during the entire comet mission between August 2014 and September 2016 to an inverse coma model.
We retrieve surface emissions on a cometary shape with 3996 triangular elements for 50 separated time intervals.
For each gas we derive systematic error bounds and report the temporal evolution of the production, peak production, and the time-integrated total production.
We discuss the production for the two lobes of the nucleus and for the northern and southern hemispheres.
Moreover we provide a comparison of the gas production with the seasonal illumination.
\end{abstract}


\section{Introduction}\label{sec:intro}

Comet 67P/Churyumov-Gerasimenko (67P/C-G) was the main rendezvous target of the European Space Agency {\it Rosetta} mission during one apparition with perihelion occurring on August 13th 2015.
The nucleus of comets consists of a mixture of frozen volatiles and of refractory components including solid organic matter (\cite{Bardyn2017}, \cite{Fray2016}).
{\it Rosetta} provided continuous {\it in-situ} and remote sensing observational data from inside the cometary coma for more than two years, see \cite{Altwegg2019} and \cite{Keller2020}.
The main tools on the {\it Rosetta} spacecraft for examining gas and dust included the instruments ROSINA (Rosetta Orbiter Spectrometer for Ion and Neutral Analysis, \cite{Balsiger2007}), VIRTIS (Visible and InfraRed Thermal Imaging Spectrometer, \cite{Coradini2007}), MIRO (Microwave Instrument for the Rosetta Orbiter, \cite{Gulkis2007}), ALICE (an ultraviolet imaging spectrograph, \cite{Stern2007}), GIADA (The Grain Impact Analyser and Dust Accumulator, \cite{Colangeli2007}), COSIMA (COmetary Secondary Ion Mass Analyzer, \cite{Kissel2007}), and OSIRIS (Optical, Spectroscopic, and Infrared Remote Imaging System, \cite{Keller2007}).
The density and composition of the cometary gas has been probed {\it in-situ} by ROSINA based on the three sensors COPS (COmet Pressure Sensor), DFMS (Double Focusing Mass Spectrometer), and RTOF (Reflectron-type Time Of Flight).
In addition, the coma was analyzed with the remote sensing instruments MIRO \citep{Biver2019} and VIRTIS \citep{Bockelee-morvan2016}.
\cite{Hansen2016} compiled the H$_2$O production obtained from these various instruments and others.

An important quantity to study is the emission rate from the nucleus.
To establish a relation between {\it in-situ} gas densities and the (sub)surface sublimation (and thus the ice composition) requires a suitable model to trace the gas release from the nucleus into the coma.
The analytical model suggested by \cite{Haser1957} provides a first estimate for the coma density under the assumption of a uniformly gas emitting spherical nucleus.
More complex coma models are described by \cite{Tenishev2008}, \cite{Fougere2013}, \cite{Bieler2015}, and \cite{Combi2020}.
These models are based on gas kinetic equations and have to incorporate the boundary conditions at the ice-gas interface, in addition to the solar illumination, the non-spherical shape of the nucleus, and the local surface properties.
Currently, no complete understanding of the ice-gas interface exists and most advanced coma models predict the spatial and temporal evolution of the coma solely based on the shape of the nucleus and the notion of an active surface area in conjunction with the local illumination, see for instance \cite{Keller2015a}.
For comet 67P/C-G several authors fit observational data to coma models to extract production data.
An independent extraction of the gas production is provided by the analysis of the the rotational state and the non-gravitational acceleration of the nucleus, see \cite{Kramer2019}, \cite{Kramer2019a}, \cite{Attree2019}, \cite{Mottola2020}.

The coma of comet 67P/C-G is dominated by three major gas species (H$_2$O, CO$_2$, and CO), which comprise 90\% of the total gas production (see Table~\ref{tab:prod}).
Coma densities of the major gas species are derived by  \cite{Bieler2015}, \cite{Bockelee-morvan2015}, \cite{Fink2016}, \cite{Marshall2017}, \cite{Biver2019}, \cite{Lauter2019}, and \cite{Combi2020}.
Areas of different relative abundances for these gases on the surface are analyzed by \cite{Hassig2015} and \cite{Hoang2017} using nadir mappings to an idealized spherical surface.
\cite{Fougere2016a}, \cite{Fougere2016}, and \cite{Hansen2016} fit 25 coefficients of a 4th order spherical harmonics expansion to locate gas activity.
\cite{Zakharov2018} and \cite{Marschall2017} consider illumination conditions, the latter additionally surface properties, to further constrain the surface activity.
\cite{Kramer2017} and \cite{Lauter2019} carry out a surface localization of gas production on  triangular shape models with different resolutions and reported strong correlations of enhanced surface emitters with reported dust outbreaks around perihelion by \cite{Vincent2016a}.

A detailed inventory of the cometary coma requires to look beyond the three major species.
In the absence of chemical reactions in the gas phase, all coma measurements at typical {\it Rosetta} distances from the nucleus are linked to the molecular abundances of the ices and grains on the nucleus.
The fingerprint of an extended set of minor volatiles in the coma provides insights in the formation processes of the early solar system, see \cite{Ahearn2012}.
{%
While the three major species reflect mainly the physical conditions under which comets formed (e.g. temperature and location), minor species reflect the chemical complexity of the native environment of comets.
The correlation between the sublimation of major and minor species is complex, as the minor species are most likely embedded in a matrix of major species.
Therefore minor species do not sublimate at their own sublimation temperature but will be released with their matrix.
Following the respective coma composition locally over the cometary orbit around the sun allows one to understand the mixture of species in the cometary ice and their release.
Remote sensing observations of comets are restricted to relatively short time periods due to signal strengths, geometrical limits, and availability of antennas.
In order to be able to compare comets, it is therefore important to understand the different outgassing patterns for the species over a large range of heliocentric distances. 
}
The recent observation of interstellar comets requires to establish an inventory of volatiles and production rates to detect novel signatures.
For the interstellar comet 2I/Borisov, \cite{Bodewits2020} and \cite{Cordiner2020} report a notably high abundance relative to water for CO.
For a number of different comets \cite{Biver2019}, \cite{Biver2018}, \cite{Biver2002}, \cite{Biver1999}, \cite{Enzian1999}, and \cite{Biver1997a} study the temporal evolution of production rates close to perihelion.
For comet 67P/C-G \cite{Luspay-kuti2015} report correlations of minor species with either H$_2$O or CO$_2$.
From the varying time evolution of the hemispheric gas production also among minor species \cite{Bockelee-morvan2016} conclude the existence of regional volatile-poor surface layers.
\cite{Calmonte2016} analyze sulphur containing molecules released from 67P/C-G, \cite{Rubin2018} detect several noble gases.

For the major gas species (water H$_2$O, carbon dioxide CO$_2$, and carbon monoxide CO) and minor gas species (hydrogen sulfide H$_2$S, oxygen O$_2$, ethane C$_2$H$_6$, methanol CH$_3$OH, formaldehyde H$_2$CO, methane CH$_4$, ammonia NH$_3$, hydrogen cyanide HCN, ethanol C$_2$H$_5$OH, carbonyl sulfide OCS, carbon disulfide CS$_2$) we derive the time-evolution of the production rates and emission regions.
The same DFMS data set is analyzed for a short time period between May 22nd and June 2nd 2015 in \cite{Rubin2019} and references therein.
Our analysis spans almost the whole {\it Rosetta} mission to 67P/C-G between 2014 and 2016.
Sect.~\ref{sec:data} introduces our derivation of 
surface-emission rates based on the shape model of comet 67P/C-G consisting of 3996 triangles.
We process all COPS/DFMS data for 50 separate time intervals and apply a $2\sigma$ criterion to mask data outliers.
In Sect.~\ref{sec:global} we present the temporal evolution of 14 volatiles during the spacecraft mission and obtain time-integrated productions and peak productions.
We discuss and compare our results to other reported observations and productions in Sect.~\ref{sec:comp}, followed by the conclusions in Sect.~\ref{sec:conclusions}.

\section{Data processing and model setup}\label{sec:data}

Our data analysis combines an analytical model for the expansion of a collisionless gas into space with an optimization procedure to constrain a large number of emission sources  (see \cite{Kramer2017} and \cite{Lauter2019}).
The measured  {\it in-situ} gas density is the superposition of the gas expansion from separated gas sources placed on a triangular mesh of the cometary surface.
For the gas expansion the model assumes the gas release from the cometary surface although sublimation processes occur in the sub-surface layers of the soil column, see \cite{Skorov2020}.
{To limit the number of unknowns, the analysis in Sects.~\ref{sec:global} and \ref{sec:comp} is done on a mesh with $N_\mathrm{E}=3996$ triangular faces with an average diameter of 120\,m, derived from the mesh given by \cite{Preusker2017}.
The resolution of the shape model has only a small influence on the derived production rates, with peak productions changing less than 5\% upon switching to a coarser-grained shape model (1024 faces).
}
\begin{figure*}
\includegraphics[width=0.962\textwidth]{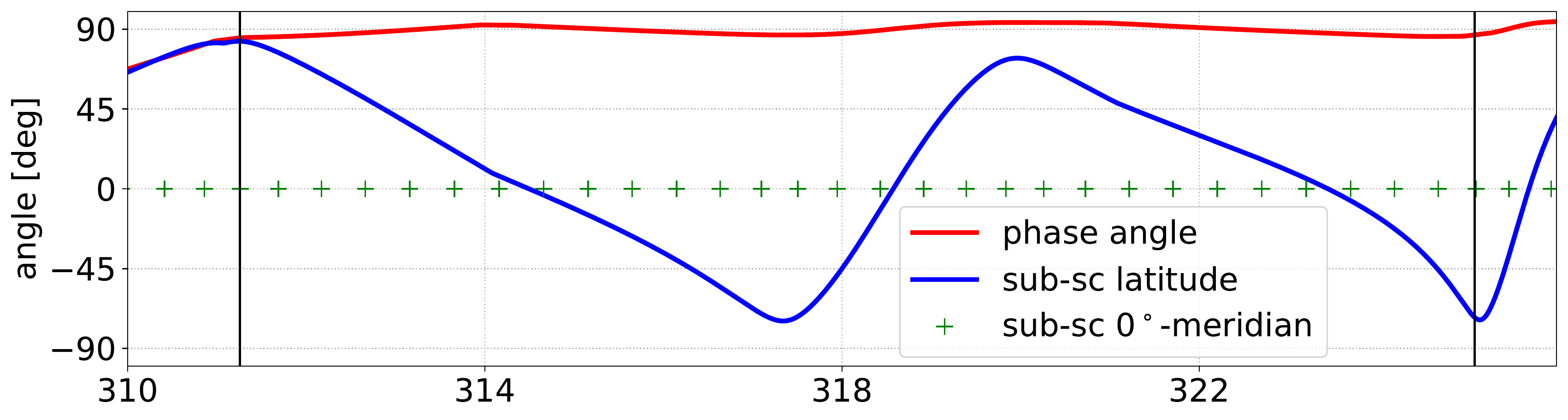}\\
\includegraphics[width=0.97\textwidth]{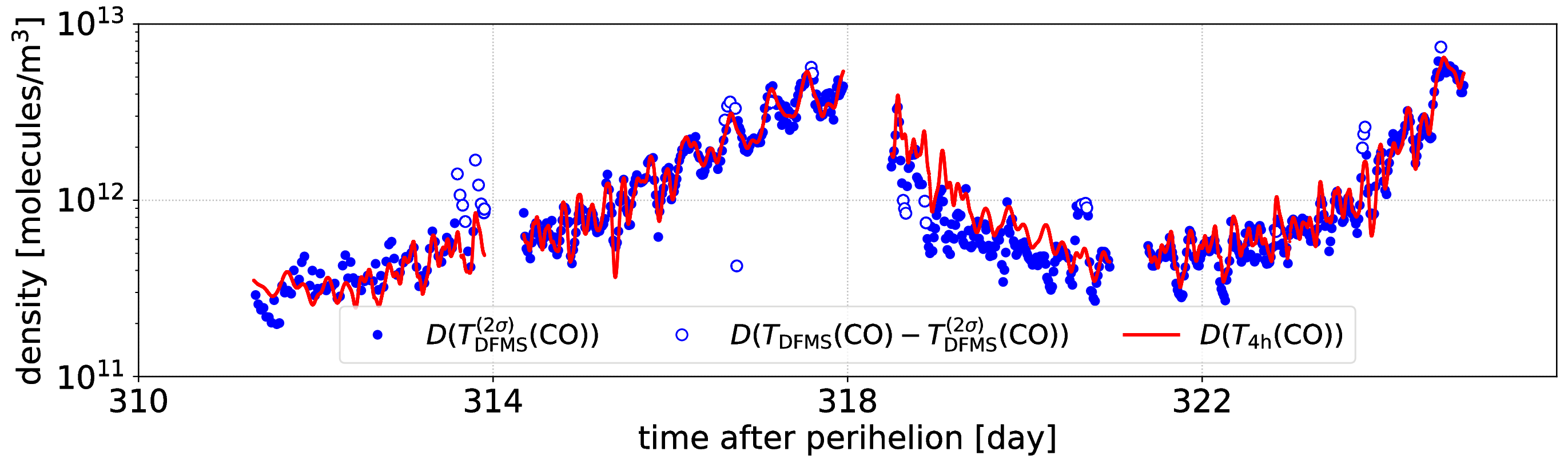}
\caption{Observing geometry and data reduction for the density probes by the spacecraft within the exemplary interval $I_{45} = (311.3\,\mathrm{d},325.1\,\mathrm{d})$.
Top panel: phase angle, sub-spacecraft latitude, and $0^\circ$-meridian crossings of the spacecraft with respect to the cometary nucleus.
Bottom panel: measured CO densities; circles (filled and open) denote COPS/DFMS densities at the times $T_{\mathrm{DFMS}}(\mathrm{CO},I_{45})$.
The filled circles marks the COPS/DFMS densities at the times $T^{(2\sigma)}_{\mathrm{DFMS}}(\mathrm{CO},I_{45})$, the open circles show discarded COPS/DFMS densities due to the $2\sigma$ criterion.
The red line denotes the linearly interpolated densities at the times
$T_{4\mathrm{h}}(\mathrm{CO},I_{45},T_{\mathrm{DFMS}})$ in Eq.~\eqref{equ:dfms}.}
\label{fig:sigma}
\end{figure*}
\cite{Gasc2017a} detail how the combined data from the two ROSINA instruments COPS and DFMS determines the {\it in-situ} gas densities of the 14 gas species
\begin{multline}\label{eq:species}
S = \{ \mathrm{H}_2\mathrm{O}, \mathrm{CO}_2{}, \mathrm{CO}{}, \mathrm{H}_2\mathrm{S}{},
\mathrm{O}_2{}, \mathrm{C}_2\mathrm{H}_6{}, \mathrm{CH}_3\mathrm{OH}{}, \mathrm{H}_2\mathrm{CO}{}, \\
\mathrm{CH}_4{}, \mathrm{NH}_3{}, \mathrm{HCN}{}, \mathrm{C}_2\mathrm{H}_5\mathrm{OH}{},
\mathrm{OCS}{}, \mathrm{CS}_2{} \}.
\end{multline}
%
% DFMS
By itself, DFMS data determines relative molecular abundances only and COPS data is required to convert relative densities to absolute ones.
We consider all measurements of the gas densities between August 1st 2014 (377\,$\mathrm{d}$ before perihelion) and September 5th 2016 (390\,$\mathrm{d}$ after perihelion).
With the convention to use negative values for days before perihelion the complete mission interval is denoted by $(-377\,\mathrm{d},390\,\mathrm{d})$.
The analysis proceeds in $N_\mathrm{I}=50$ separated subintervals
\[
I_1,I_2,..., I_{N_\mathrm{I}} \subset ( -377\,\mathrm{d},390\,\mathrm{d} ).
\]
The subintervals last between $7\,\mathrm{d}$ and $29\,\mathrm{d}$.
Within each subinterval the sub-spacecraft position samples almost the entire surface of the nucleus.
The chosen subinterval duration ensures a limited variation of the heliocentric distance $r_\mathrm{h}$ and the subsolar latitude.
This allows us to neglect seasonal changes in the sublimation rate within each subinterval.
The exemplary subinterval $I_{45} = (311.3\,\mathrm{d},325.1\,\mathrm{d})$ is shown in Fig.~\ref{fig:sigma}.
To constrain the impact of a varying phase angle between {\it Rosetta} and the nucleus and the diurnal changes in the sublimation rate we have highlighted all subintervals with {\it Rosetta} observations around an phase angle of 90$^\circ$ in section \ref{sec:global}. 
This operational orbit is sometimes referred to as terminator orbit.

The DFMS measurements are conducted less frequently than the COPS ones.
The density $\rho_{s}(t)$ of a gas species $s\in S$ is recorded at times $t$ in $T_{\mathrm{DFMS}}=T_{\mathrm{DFMS}}(s,I_j)$ within the subinterval $I_j$.
Each species and subinterval contains a distinct set of DFMS measurements
\begin{equation*}
D(T_{\mathrm{DFMS}}) = \{ ( t, \rho_{s}(t) )\; |\; t\in T_{\mathrm{DFMS}} \}.
\end{equation*}
The data set encompassing all species and subintervals contains 218,765 entries.
Only for a small number of species and subintervals no data is available.
For the exemplary species CO in the interval $I_{45}$ the data points are shown in Fig.~\ref{fig:sigma}.
Due to spacecraft maneuvers the ROSINA sensors experience standby or off mode.
To maintain a sufficient surface resolution for mapping the DFMS data to gas emitters close to the surface requires to interpolate between neighbouring DFMS data points to times when COPS data is available, as discussed by \cite{Lauter2019}.
During the entire comet mission COPS measurements of the gas density are available at about $10^6$ spacecraft positions, enumerated by the corresponding observation times from the set $T_\mathrm{COPS}$.
The number of DFMS data points is increased by a linear interpolation of the densities $\rho_{s}(t)$ to an extended set of times $T_{4\mathrm{h}}=T_{4\mathrm{h}}(s,I_j,T_{\mathrm{DFMS}})$, as described in \cite[Eq. (1)]{Lauter2019}.
The set $T_{4\mathrm{h}}$ consists of only those times $t\in T_\mathrm{COPS}\cap I_j$ that are enclosed in a time interval $(t_l,t_r)$ with a length of at most 4\,$\mathrm{h}$ and with DFMS times $t_l,t_r \in T_{\mathrm{DFMS}}$.
For each gas species $s\in S$ in each time interval $I_j$ linear interpolation yields the extended data set
\begin{equation}\label{equ:dfms}
D(T_{4\mathrm{h}}) = \{ ( t, \rho_{s}(t) ) \; |\; t\in T_{4\mathrm{h}} \}.
\end{equation}
The number of data points depends on the species and the interval and varies from 4,752 points for CH$_3$OH to 19,358 points for H$_2$O.
Fig.~\ref{fig:sigma} shows the extended data set for CO in the interval $I_{45}$.

{%
According to Eqs.~(1) and (5) in \cite{Kramer2017} the analytic gas model describes the density
\[
\rho_{s,j}(\vec{x}_\mathrm{sc}) = \sum_{i=1}^{N_\mathrm{E}} \mathrm{occ}_i(\vec{x}_\mathrm{sc}) \rho_{s,i,j}(\vec{x}_\mathrm{sc})
\]
at the spacecraft position $\vec{x}_\mathrm{sc}$ as a superposition of contributions $\rho_{s,i,j}$ for each species $s$ in each interval $I_j$ with the occultation function $\mathrm{occ}_i$.
The gas density of \cite{Narasimha1962} reads
\[
\rho_{s,i,j}(\vec{x}_\mathrm{sc})
= \frac{U_0^2}{u_{s,0}}\frac{\cos\theta}{\pi \vec{r}^2}
|E_i|\dot\rho_{s,i,j} \exp(-U_0^2 \sin^2\theta)
\]
for a point source of a collisionless gas on a surface element $E_i$ with its center $\vec{b}_i$ (local position vector $\vec{r}=\vec{x}_\mathrm{sc}-\vec{b}_i$), the outward normal vector $\vec{\nu}_i$ and the angle $\theta$ such that
$\cos\theta = \vec{r} / |\vec{r}| \cdot \vec{\nu}_i$.
The ratio between the normal component $u_{s,0}$ and the lateral one of the outflow velocity is the parameter $U_0$ which is taken to $U_0=3$ as in \cite{Lauter2019}.
The surface emission-rate $\dot \rho_{s,i,j}$ is the result of a parameter fit based on system
\begin{equation}\label{eq:optim}
\rho_{s,i}(\vec{x}_\mathrm{sc}(t)) = \rho
\end{equation}
using the measurements $(t,\rho)$ in the data set $D(T_{4\mathrm{h}})$.
Eq.~(7) in \cite{Kramer2017} yields the relative $l^2$-error for the fit.
$\dot \rho_{s,i,j}$ is constant within the entire subinterval $I_j$, takes the value of the diurnally averaged gas production and depends on the outflow velocity $u_{s,0}$.
}
The velocity $u_{s,0}$ is a function of heliocentric distance since we use the parameterization for water given by \cite{Hansen2016}, which resembles the expansion velocity derived from molecular lines by \cite{Biver2019} of different species.
In particular within each interval the velocity for the molecules of all species is assumed to be the same.
Another option is to consider decoupled gases as in \cite{Lauter2019} where $u_{s,0}$ is scaled by the square root of the molecular mass relative to water.
For the latter case the density values have to be re-scaled for each species by a constant factor which varies from $\approx{}0.5$ for CS$_2$ to $\approx{}1.1$ for CH$_4$.

The data set in Eq.~\eqref{equ:dfms} was the basis for \cite{Lauter2019} to analyze major gas species.
To extend the previous analysis of COPS/DFMS data to 14 species we refine the data processing.
{Several minor species are affected by additional noise due to small concentrations, resulting in significant fit errors or lack of convergence for some intervals.}
We detect outliers in the data set by applying a $2\sigma$ criterion for the $l^2$-error functional and discard any data outside this bound. 
The standard deviation is obtained for the difference between the evaluation of our coma model density $m_{s}(t)$ at the spacecraft distance $d_{\mathrm{sc}}(t)$ and the times $t\in T_{\mathrm{DFMS}}$, and the measured data.
The rational behind this selection is to discard sudden drops and outbursts in the data from the overall repetitive outgassing behaviour of comet 67P/C-G.
This is also reflected in diurnally repeating dust pattern, see \cite{Kramer2015a,Kramer2018}.
Formally, the squared standard deviation $\sigma_{s,j}$ for the distance-weighted density $d_{\mathrm{sc}}^2 \rho_{s}$ is given by
\[
\sigma^2_{s,j} = \frac{1}{|T_{\mathrm{DFMS}} |} \sum_{t \in T_{\mathrm{DFMS}}}
\left( d_{\mathrm{sc}}^2(t)\, |\rho_{s}(t)-m_{s}(t)| \right)^2
\]
in the interval $I_j$. 
We define a reduced set of times satisfying the $2\sigma$ criterion by
\[
T^{(2\sigma)}_{\mathrm{DFMS}}
= \{ t \in T_{\mathrm{DFMS}}\; | \; d^2_{\mathrm{sc}}|\rho_{s}(t)-m_{s}(t)| < 2\sigma_{s,j} \}.
\]
This yields the reduced density set
$D(T^{(2\sigma)}_{\mathrm{DFMS}}) = \{ ( t, \rho_{s}(t) )\; |\; t\in T^{(2\sigma)}_{\mathrm{DFMS}} \}$
for DFMS data, for which we re-run the model fit.
For all species and intervals together these data sets contain 187,068 entries corresponding to 14\% less data.
The filled circles in Fig.~\ref{fig:sigma} represent this reduced data set in the exemplary interval $I_{45}$.
The $4\mathrm{h}$ criterion including linear interpolation for the densities as above yields the increased number of time points $T^{(2\sigma)}_{4\mathrm{h}}=T^{(2\sigma)}_{4\mathrm{h}}(s,I_j,T^{(2\sigma)}_{\mathrm{DFMS}})$ with the extended data set
\begin{equation}\label{equ:sigmadata}
D(T^{(2\sigma)}_{4\mathrm{h}}) = \{ ( t, \rho_{s}(t) )\; |\; t\in T^{(2\sigma)}_{4\mathrm{h}} \}
\end{equation}
for each gas species $s$ in each interval $I_j$.
The number of data points ranges from 4,443 points for CO to 19,212 points for H$_2$O.
The comparison with the number of points from Eq.~\eqref{equ:dfms} shows that the $2\sigma$ criterion data set does only remove few points.
The surface emission rates $\dot\rho_{s,i,j}$ derived from the data complying with Eq.~\eqref{equ:sigmadata} are used for the {subsequent} analysis in Sect.~\ref{sec:global}.
{For each species the uncertainty of the retrieved gas production due to the fit error (in Eq.~\eqref{eq:optim} with respect to the measured densities) is estimated by comparing production changes and fit errors for two separated model runs realized with $2\sigma$ and $8\sigma$ data.
For the $2\sigma$ data, the average fit error of about $20$\% results in production errors of about $7$\%.
In Sect.~\ref{sec:global} we detail how the fit uncertainty contributes to the overall uncertainty estimation.
}

The main computational effort for the inverse coma model is located in two code sections.
First, at each considered spacecraft position (associated with a measured density) the evaluation of the analytical model requires the complete list of directly visible surface elements.
Second the numerical solution of the parameter fit {in Eq.~\eqref{eq:optim}} is based on a standard singular value decomposition.
Each of the 14 gases and each of the 50 time intervals $I_1, I_2, ..., I_{N_\mathrm{I}}$ is assigned to one MPI (message passing interface) process.
Within each process both code sections are executed by 9 parallel (OpenMP) threads.
For one gas in one interval the analysis takes 90\,min which yields approximately 100 node hours for the complete analysis on the HLRN-IV supercomputer (96 cores per node).

\section{Evolution of the gas production}\label{sec:global}

\begin{table}
\begin{tabular}{lllc}
$s$ & $P_s\, (\mathrm{kg})$ & $P_s\, (\text{molecules})$ & $P_s/P_{\mathrm{H}_2\mathrm{O}}$ \\ \hline
H$_2$O & $[ 4.0 \pm 0.6 ] \times 10^{9}$ & $[ 1.3 \pm 0.2 ] \times 10^{35}$ & 1 \\
CO$_2$ & $[ 7.2 \pm 1.8 ] \times 10^{8}$ & $[ 9.8 \pm 2.5 ] \times 10^{33}$ & $7\times 10^{-2}$ \\
CO & $[ 1.9 \pm 0.4 ] \times 10^{8}$ & $[ 4.0 \pm 0.8 ] \times 10^{33}$ & $3\times 10^{-2}$ \\
H$_2$S & $[ 1.3 \pm 0.4 ] \times 10^{8}$ & $[ 2.3 \pm 0.8 ] \times 10^{33}$ & $2\times 10^{-2}$ \\
O$_2$ & $[ 1.6 \pm 0.3 ] \times 10^{8}$ & $[ 3.0 \pm 0.5 ] \times 10^{33}$ & $2\times 10^{-2}$ \\
C$_2$H$_6$ & $[ 5.5 \pm 1.2 ] \times 10^{7}$ & $[ 1.1 \pm 0.2 ] \times 10^{33}$ & $8\times 10^{-3}$ \\
CH$_3$OH & $[ 3.7 \pm 0.8 ] \times 10^{7}$ & $[ 7.0 \pm 1.4 ] \times 10^{32}$ & $5\times 10^{-3}$ \\
H$_2$CO & $[ 3.1 \pm 0.6 ] \times 10^{7}$ & $[ 6.1 \pm 1.3 ] \times 10^{32}$ & $5\times 10^{-3}$ \\
CH$_4$ & $[ 1.5 \pm 0.3 ] \times 10^{7}$ & $[ 5.6 \pm 1.2 ] \times 10^{32}$ & $4\times 10^{-3}$ \\
NH$_3$ & $[ 1.5 \pm 0.4 ] \times 10^{7}$ & $[ 5.3 \pm 1.4 ] \times 10^{32}$ & $4\times 10^{-3}$ \\
HCN & $[ 1.1 \pm 0.2 ] \times 10^{7}$ & $[ 2.4 \pm 0.5 ] \times 10^{32}$ & $2\times 10^{-3}$ \\
C$_2$H$_5$OH & $[ 1.3 \pm 0.3 ] \times 10^{7}$ & $[ 1.7 \pm 0.4 ] \times 10^{32}$ & $1\times 10^{-3}$ \\
OCS & $[ 9.6 \pm 2.1 ] \times 10^{6}$ & $[ 9.7 \pm 2.1 ] \times 10^{31}$ & $7\times 10^{-4}$ \\
CS$_2$ & $[ 3.2 \pm 0.6 ] \times 10^{6}$ & $[ 2.5 \pm 0.5 ] \times 10^{31}$ & $2\times 10^{-4}$
\end{tabular}%
\caption{Time-integrated productions $P_s$ for all species $s$ in Eq.~\eqref{eq:species} for the complete {\it Rosetta} mission time ranging from $-377\,\mathrm{d}$ before to $390\,\mathrm{d}$ after perihelion. The production relative to water is based on the number of molecules.
{Data uncertainties are discussed in Sect.~\ref{sec:global}.}
}
\label{tab:prod}
\end{table}
\begin{table}
\begin{tabular}{llc}
$s$ & $\max Q_{s}\, (\text{molecules/s})$ & $\max Q_{s}/\max Q_{\mathrm{H}_2\mathrm{O}}$ \\ \hline
H$_2$O & $[ 1.85\pm 0.03 ] \times 10^{28}$ & 1 \\
CO$_2$ & $[ 1.58 \pm 0.07 ] \times 10^{27}$ & $9 \times 10^{-2}$ \\
CO & $[ 5.9 \pm 1.7 ] \times 10^{26}$ & $3 \times 10^{-2}$ \\
H$_2$S & $[ 4.4 \pm 0.5 ] \times 10^{26}$ & $2 \times 10^{-2}$ \\
O$_2$ & $[ 3.6 \pm 0.4 ] \times 10^{26}$ & $2 \times 10^{-2}$ \\
C$_2$H$_6$ & $[ 1.58 \pm 0.05 ] \times 10^{26}$ & $9 \times 10^{-3}$ \\
CH$_3$OH & $[ 1.14 \pm 0.05 ] \times 10^{26}$ & $6 \times 10^{-3}$ \\
H$_2$CO & $[ 9.7 \pm 0.9 ] \times 10^{25}$ & $5 \times 10^{-3}$ \\
CH$_4$ & $[ 8.2 \pm 1.2 ] \times 10^{25}$ & $4 \times 10^{-3}$ \\
NH$_3$ & $[ 1.0 \pm 0.1 ] \times 10^{26}$ & $5 \times 10^{-3}$ \\
HCN & $[ 3.7 \pm 0.3 ] \times 10^{25}$ & $2 \times 10^{-3}$ \\
C$_2$H$_5$OH & $[ 3.0 \pm 0.2 ] \times 10^{25}$ & $2 \times 10^{-3}$ \\
OCS & $[ 1.58 \pm 0.08 ] \times 10^{25}$ & $9 \times 10^{-4}$ \\
CS$_2$ & $[ 4.56 \pm 0.07 ] \times 10^{24}$ & $2 \times 10^{-4}$
\end{tabular}%
\caption{Peak productions $\max Q_{s}=\max_j Q_{s,j}$ for all species $s$ in Eq.~\eqref{eq:species}. For each species $s$ the maximum value appears in one interval $I_{25}$ ranging from $17\,\mathrm{d}$ to $27\,\mathrm{d}$ after perihelion. The abundance relative to water is evaluated for interval $I_{25}$. 
{Data uncertainties are discussed in Sect.~\ref{sec:global}.}
}
\label{tab:peak}
\end{table}
Based on the surface emission rates in Sect.~\ref{sec:data} we evaluate the time-integrated productions $P_s$ in Table~\ref{tab:prod} and peak productions $\max_j Q_{s,j}$ in Table~ \ref{tab:peak}.
The given values are affected by several systematic uncertainties.
\cite{Rubin2019} estimate 30\% for the uncertainty of relative abundances (DFMS data) including the effects of sensitivity calibration, detector gain, and fitting errors where applicable.
Our method introduces further uncertainties with respect to {the fit error (see 
Sect.~\ref{sec:data}) and} a partially reduced surface coverage of the spacecraft trajectory.
Time intervals suffer from limited surface coverage and thus encompass areas with an unassigned production rate (not-seen surface elements).
To constrain the unknown surface production originating from these areas we provide a lower and an upper estimate of the production in the time interval $I_j$.
The lower bound is given by setting the unknown surface emission rate to zero, the upper bound is provided by the maximum value of the production rate from the same surface elements within the neighbouring intervals $I_{j-1}$ and $I_{j+1}$.
If a lack of surface coverage results in an unassigned production value on one element  for $I_{j-1}$, $I_{j}$, and $I_{j+1}$, then the production is set to zero.
These estimates {for the gas production with respect to limited surface coverage} simplify the analysis compared to \cite{Lauter2019}, where we additionally considered a linear interpolation across additional intervals.
{%
Our overall uncertainty estimation given in Tabs.~\ref{tab:prod}, \ref{tab:peak} and the figures assumes uncorrelated uncertainties for
the fit error and the limited surface coverage.
}
This applies to the time-integrated productions $P_s$ and in the peak
productions $\max_j Q_{s,j}$, too.

\begin{figure*}
\includegraphics[width=0.45\textwidth]{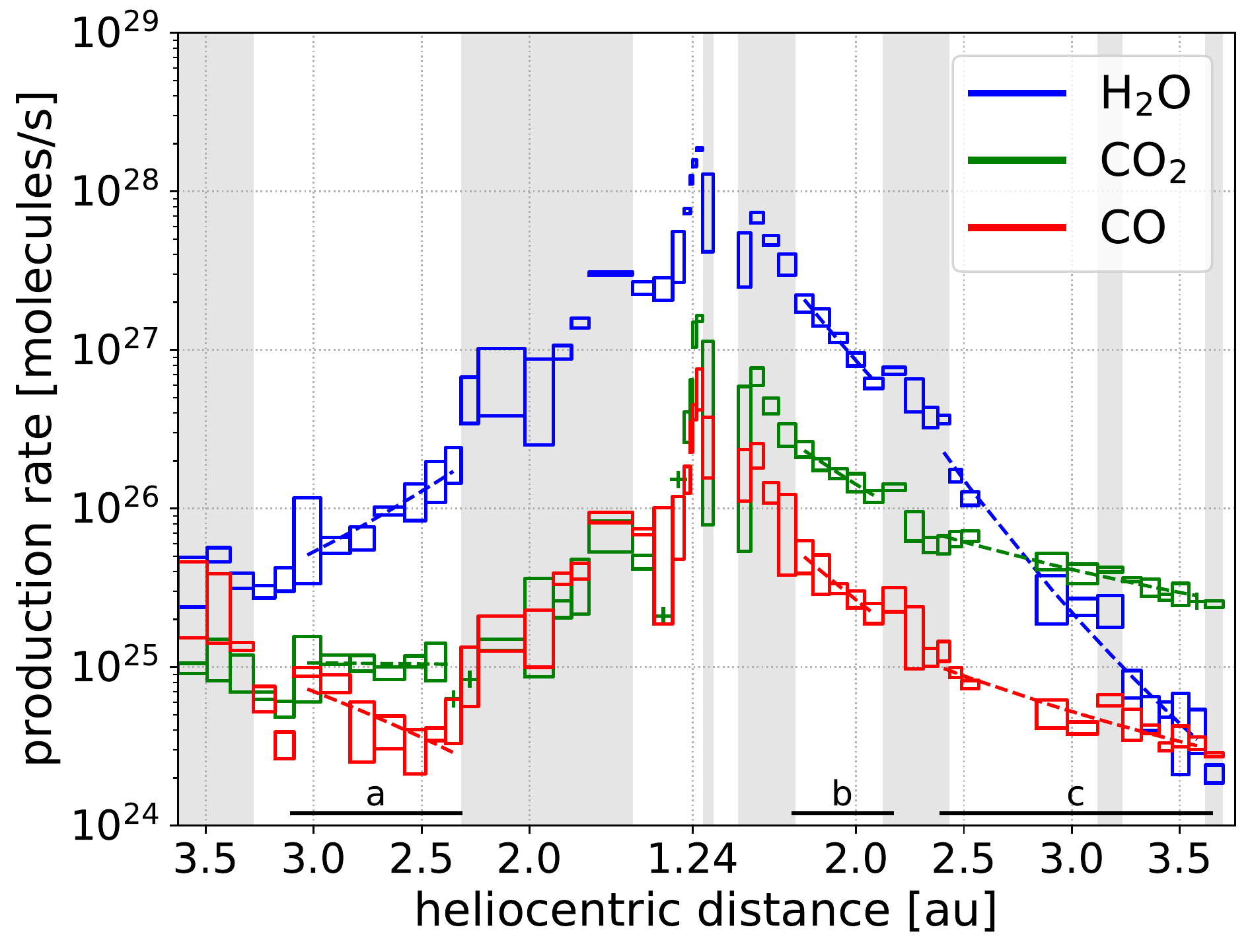}
\includegraphics[width=0.45\textwidth]{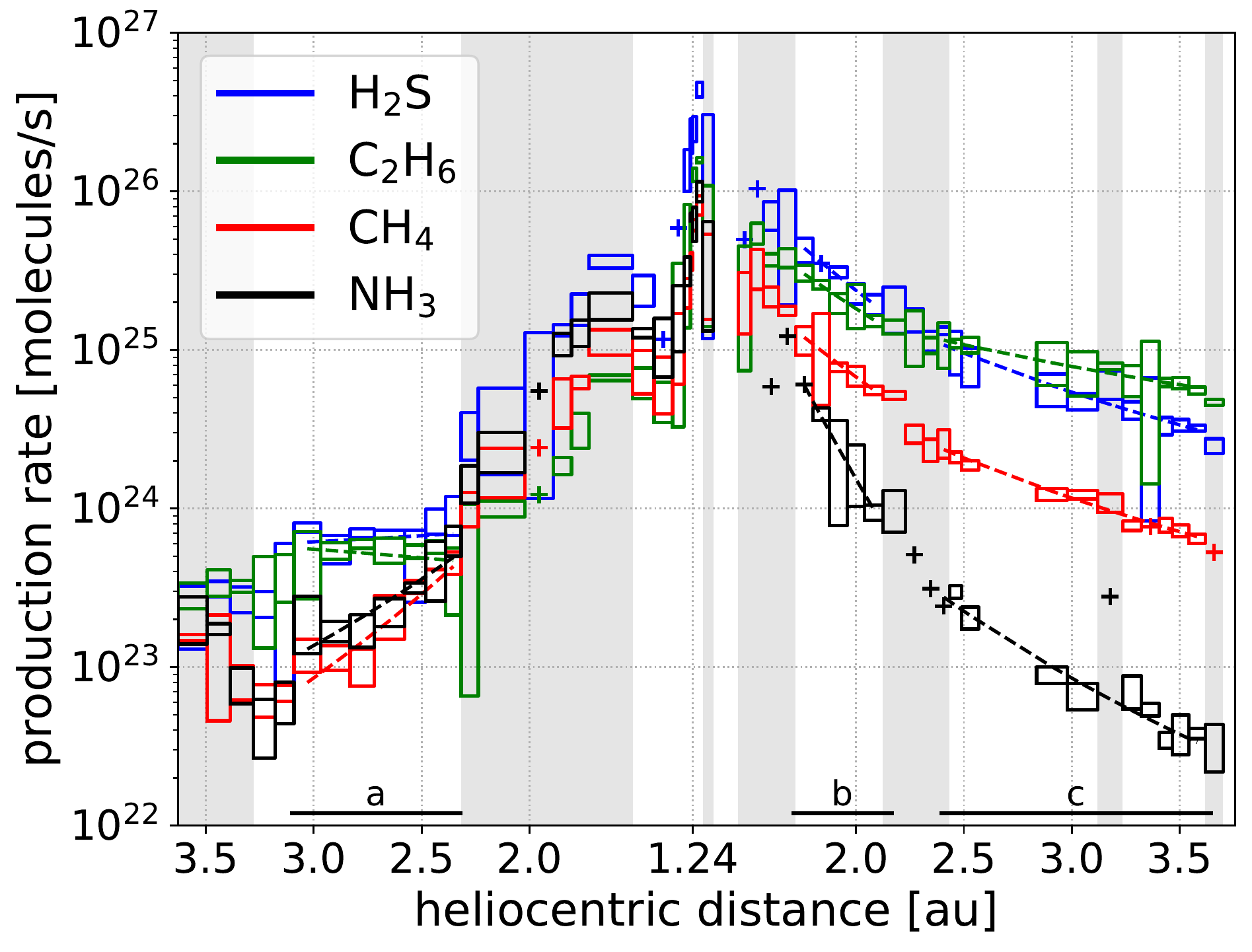}\\
\includegraphics[width=0.45\textwidth]{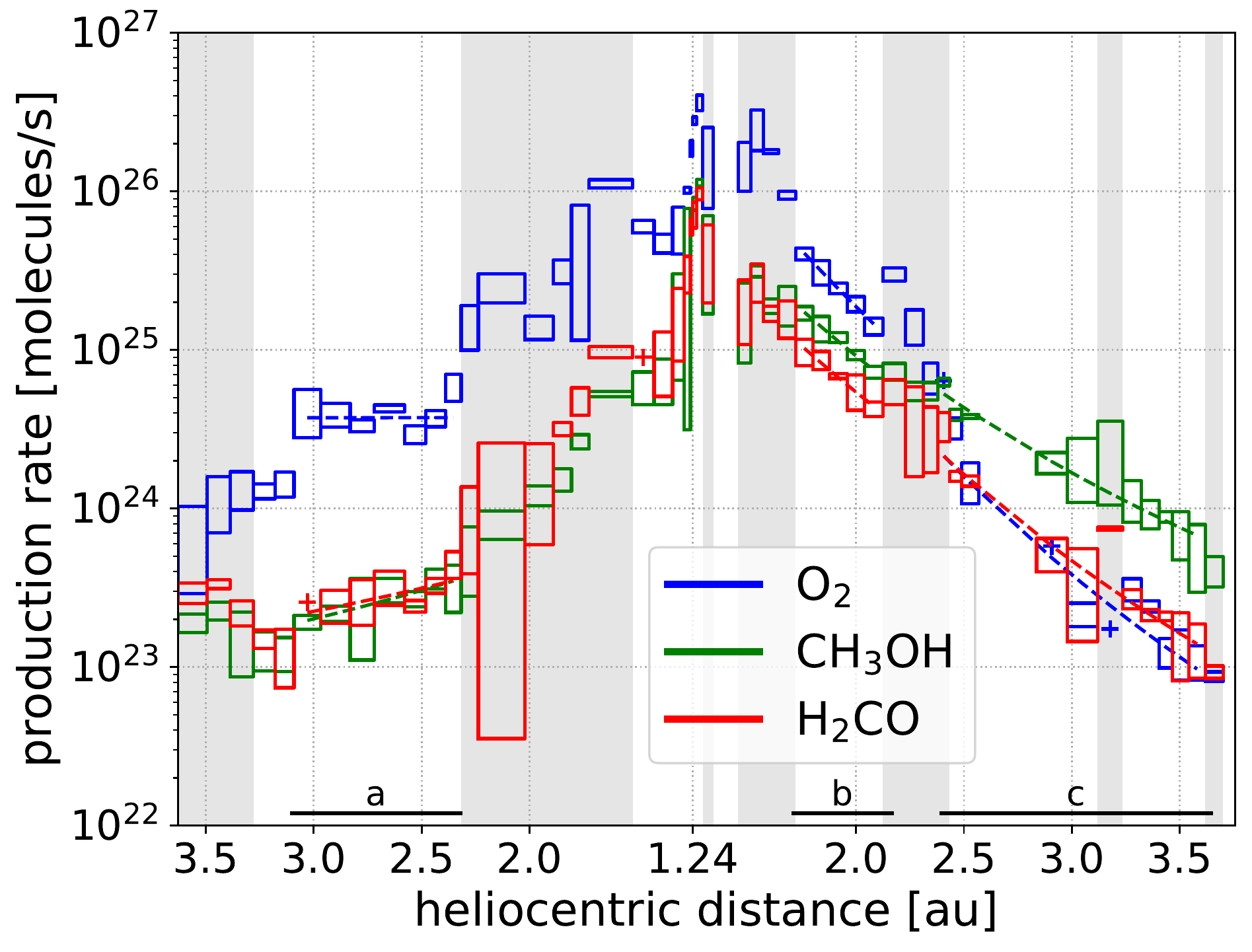}
\includegraphics[width=0.45\textwidth]{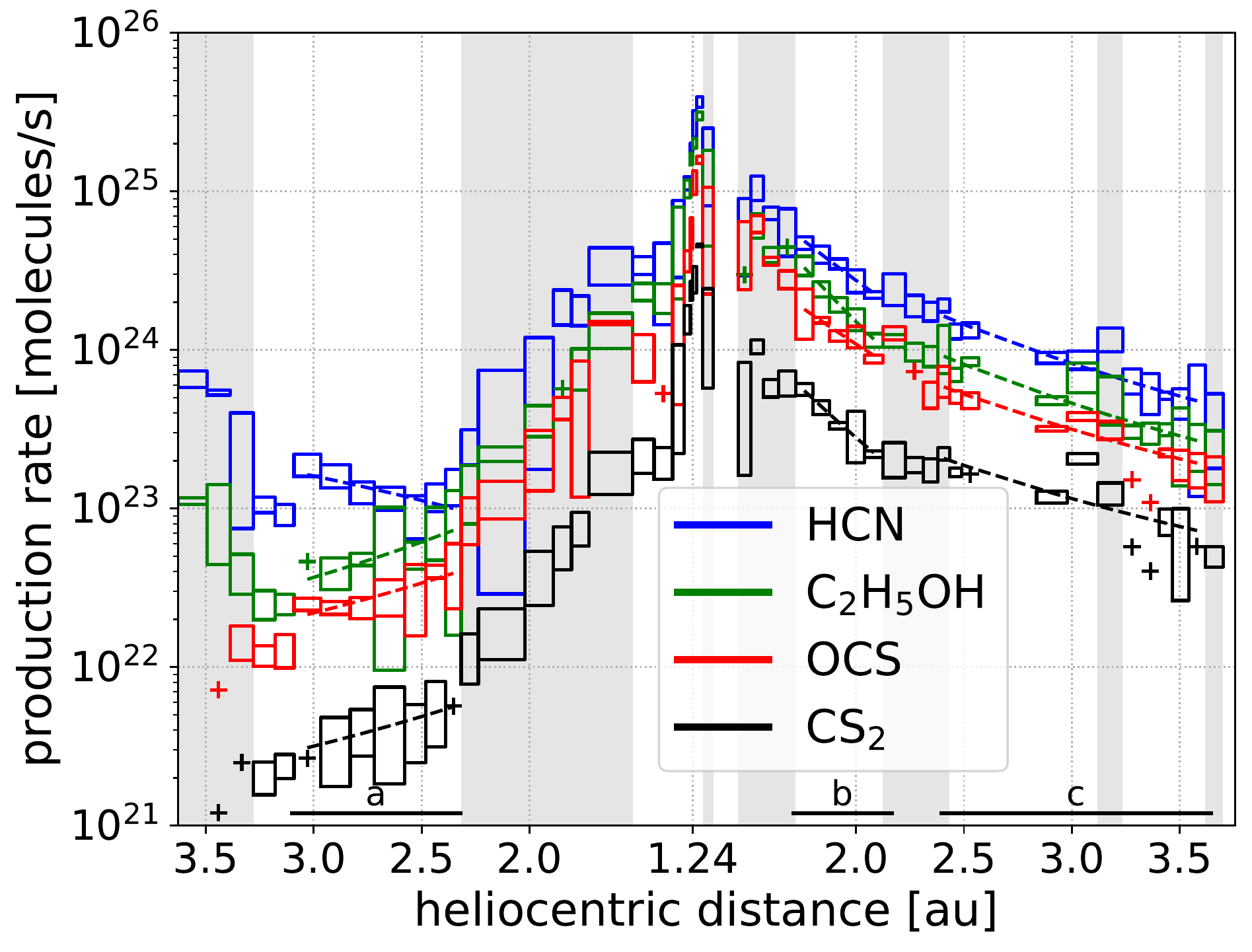}
\caption{Production rates $Q_{s,j}$ for all species $s$ in Eq.~\eqref{eq:species} as a function of heliocentric distance $r_\mathrm{h}$ of comet 67P/C-G. The boxes indicate the uncertainties {discussed in Sect.~\ref{sec:global}. Without available uncertainty estimation a + marker is used.}
Grey colored intervals denote phase angles differing from $90^\circ$, indicating non-terminator orbits. The dashed lines represent the best-fit power law in the distinct intervals $I_\mathrm{a}$, $I_\mathrm{b}$, and $I_\mathrm{c}$, indicated by the horizontal bars and tabulated in Table~\ref{tab:intervals}.}
\label{fig:gastrees}
\end{figure*}
The temporal evolution of all 14 production rates $Q_{s,j}$ is shown in Fig.~\ref{fig:gastrees}.
To further reduce sampling errors, we have preferentially chosen intervals with terminator orbits.
Terminator orbits encompass typically morning and evening illumination conditions and match with the assumption of mainly observing gas emissions representing diurnally averaged production rates.
Intervals deviating from terminator orbits are marked in grey to reveal possible systematic errors related to a varying phase angle, in particular to a more illuminated nucleus.
Almost all, except two, grey intervals correspond to average phase angles smaller than $90^\circ$.

\begin{table}
\begin{tabular}{cccc}
 & interval & $r_\mathrm{h}\,(\mathrm{au})$ & month \\ \hline
$I_\mathrm{a}$ & $(-290\,\mathrm{d},-180\,\mathrm{d})$ &
3.1\,-\,2.3 &  11/2014\,-\,02/2015 \\
$I_\mathrm{b}$ & $(100\,\mathrm{d},160\,\mathrm{d})$ &
1.7\,-\,2.2 & 11/2015\,-\,01/2016 \\
$I_\mathrm{c}$ & $(190\,\mathrm{d},380\,\mathrm{d})$ & 
2.4\,-\,3.6 & 02/2016\,-\,08/2016
\end{tabular}
\caption{Definitions for the time intervals $I_\mathrm{a}$, $I_\mathrm{b}$, and $I_\mathrm{c}$ with respect to days after perihelion, heliocentric distances $r_\mathrm{h}$, and months.}
\label{tab:intervals}
\end{table}
\begin{table}
\begin{tabular}{lcccc}
& in $I_\mathrm{a}$ & in $I_\mathrm{b}$ & in $I_\mathrm{c}$ & \\
$s$ & 3.1\,-\,2.3\,au & 1.7\,-\,2.2\,au & 2.4\,-\,3.6\,au &  group \\
\hline
H$_2$O & -5.3 & -6.5 & -9.5 & H$_2$O \\
CO$_2$ & 0.1 & -3.6 & -2.0 & CO$_2$ \\
CO & 4.0 & -4.6 & -2.6 & CO$_2$ \\
H$_2$S & -0.5 & -4.5 & -2.8 & CO$_2$ \\
O$_2$ & 0.0 & -5.7 & -7.1 & H$_2$O \\
C$_2$H$_6$ & 0.8 & -3.7 & -1.6 & CO$_2$ \\
CH$_3$OH & -2.5 & -4.7 & -4.7 & H$_2$O \\
H$_2$CO & -2.0 & -4.7 & -6.2 & H$_2$O \\
CH$_4$ & -7.4 & -4.3 & -2.9 & CO$_2$ \\
NH$_3$ & -5.8 & -10.1 & -4.8 & H$_2$O \\
HCN & 2.1 & -4.2 & -2.8 & CO$_2$ \\
C$_2$H$_5$OH & -3.1 & -5.8 & -2.8 & CO$_2$ \\
OCS & -2.6 & -3.7 & -2.5 & CO$_2$ \\
CS$_2$ & -2.6 & -5.0 & -2.4 & CO$_2$
\end{tabular}
\caption{Exponent $\alpha$ of the fitted power law $r_\mathrm{h}^{\alpha}\sim Q_{s,j}$ separate for each of the intervals $I_\mathrm{a}$, $I_\mathrm{b}$, and $I_\mathrm{c}$ in Table~\ref{tab:intervals}. The outbound exponents for $I_\mathrm{c}$ define the CO$_2$ group and H$_2$O group, see Sect.~\ref{sec:global}.}
\label{tab:powerfit}
\end{table}
\begin{table}
\begin{tabular}{lcccc}
$s$ & $A_\mathrm{N}$ & $A_\mathrm{S}$ & small lobe & big lobe \\ \hline
CO$_2$ & 3.7 & -1.4 & -0.2 & 0.3 \\
CO & 6.0 & 5.0 & 3.7 & 4.2 \\
H$_2$S & -0.2 & 0.8 & -0.5 & -0.5 \\
O$_2$ & 0.2 & 0.6 & 0.2 & -0.1 \\
C$_2$H$_6$ & 0.1 & 1.9 & 0.6 & 0.9 \\
HCN & 3.5 & 3.3 & 2.4 & 1.9
\end{tabular}
\caption{Exponent $\alpha$ of the fitted power law $r_\mathrm{h}^{\alpha}\sim Q_{s,j}$ within the inbound interval $I_\mathrm{a}$ with respect to areas of the cometary surface: $A_\mathrm{N}$ -- northern hemisphere, $A_\mathrm{S}$ -- southern hemisphere, small lobe, and big lobe.}
\label{tab:localpower}
\end{table}
The global production curve agrees with the analysis of the non-gravitational acceleration \citep{Kramer2019a} and the change of the rotation axis \citep{Kramer2019}.
We discern distinct patterns in the evolution of the production rates.
Despite the overall increasing solar illumination in the inbound orbital arc between $-290\,\mathrm{d}$ and $-180\,\mathrm{d}$ (heliocentric distances between 3.1\,au and 2.3\,au), the gas production for some gases stagnates or even decreases.
For all species the production increases toward perihelion and culminates in a pronounced peak in our interval $I_{25}$ ranging from 17\,d to 27\,d after perihelion, {see Table.~\ref{tab:peak}}.
{
Further subdivision or interleaving of our subintervals is not possible without limiting the surface coverage due to the characteristics of the subspacecraft latitude.
Based on COPS/DFMS data alone it is not possible to further constrain the day of peak production.
An independent analysis based on the non-gravitational acceleration of the nucleus 
(Fig.~(3) in \cite{Kramer2019a}) puts the maximum production around 0-20 days after perihelion.
}
The distinct peak around perihelion does not allow one to fit all observations to a single power law.
For outbound heliocentric distances exceeding $\approx 2.5\,\mathrm{au}$ we find two groups of gases with markedly different production decreases.

\begin{figure*}
\includegraphics[width=0.65\textwidth]{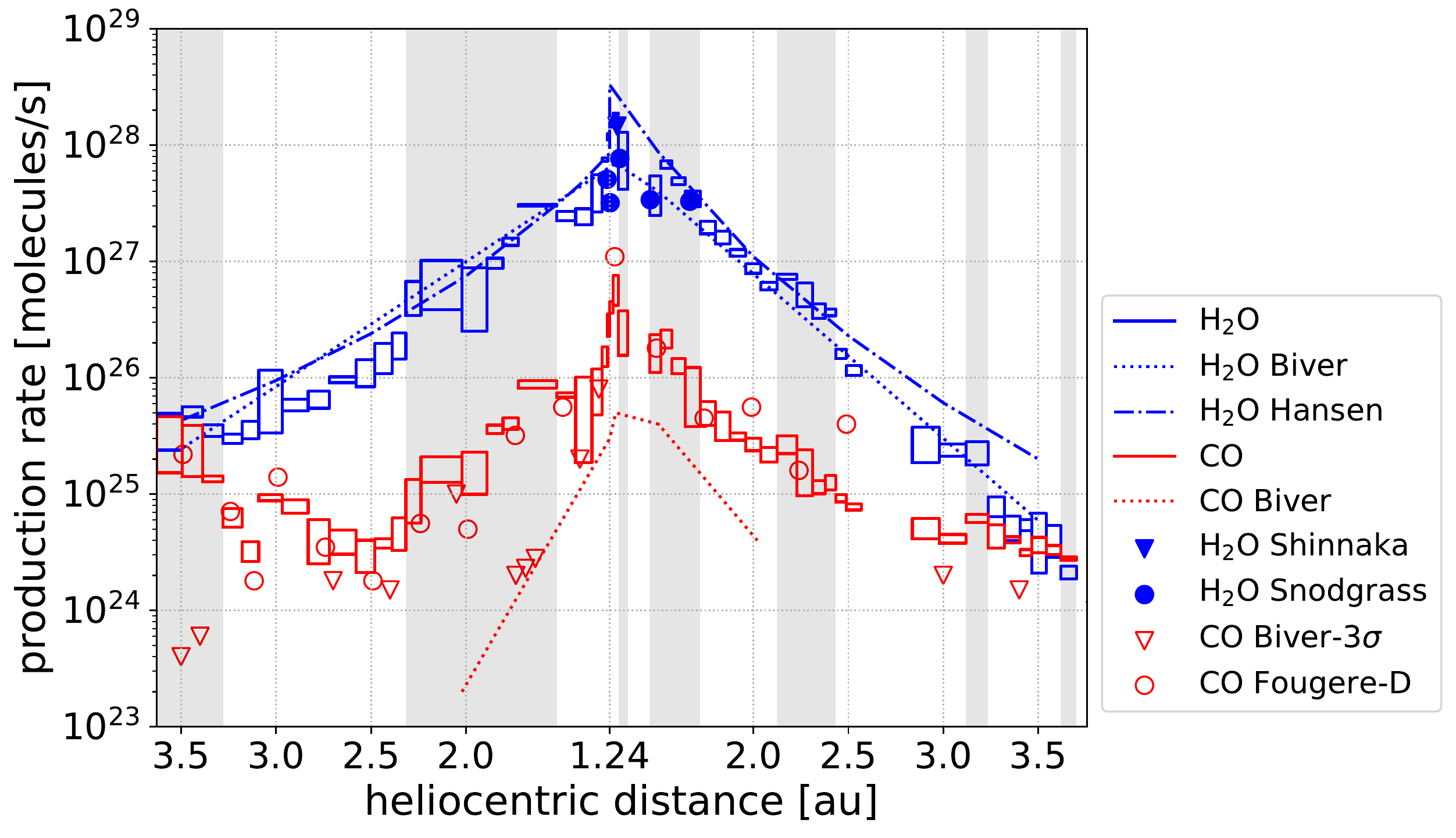}\\
\includegraphics[width=0.45\textwidth]{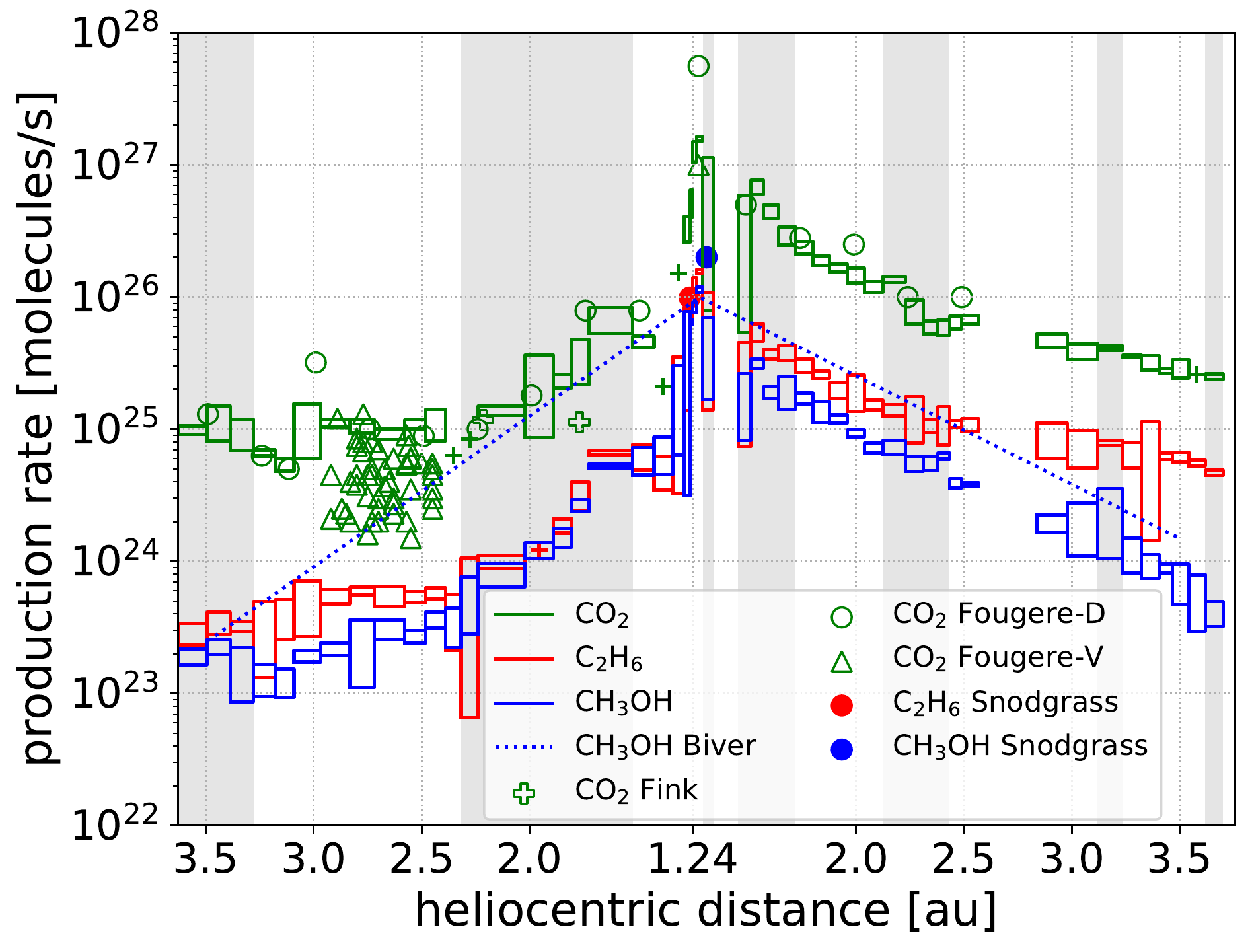}
\includegraphics[width=0.45\textwidth]{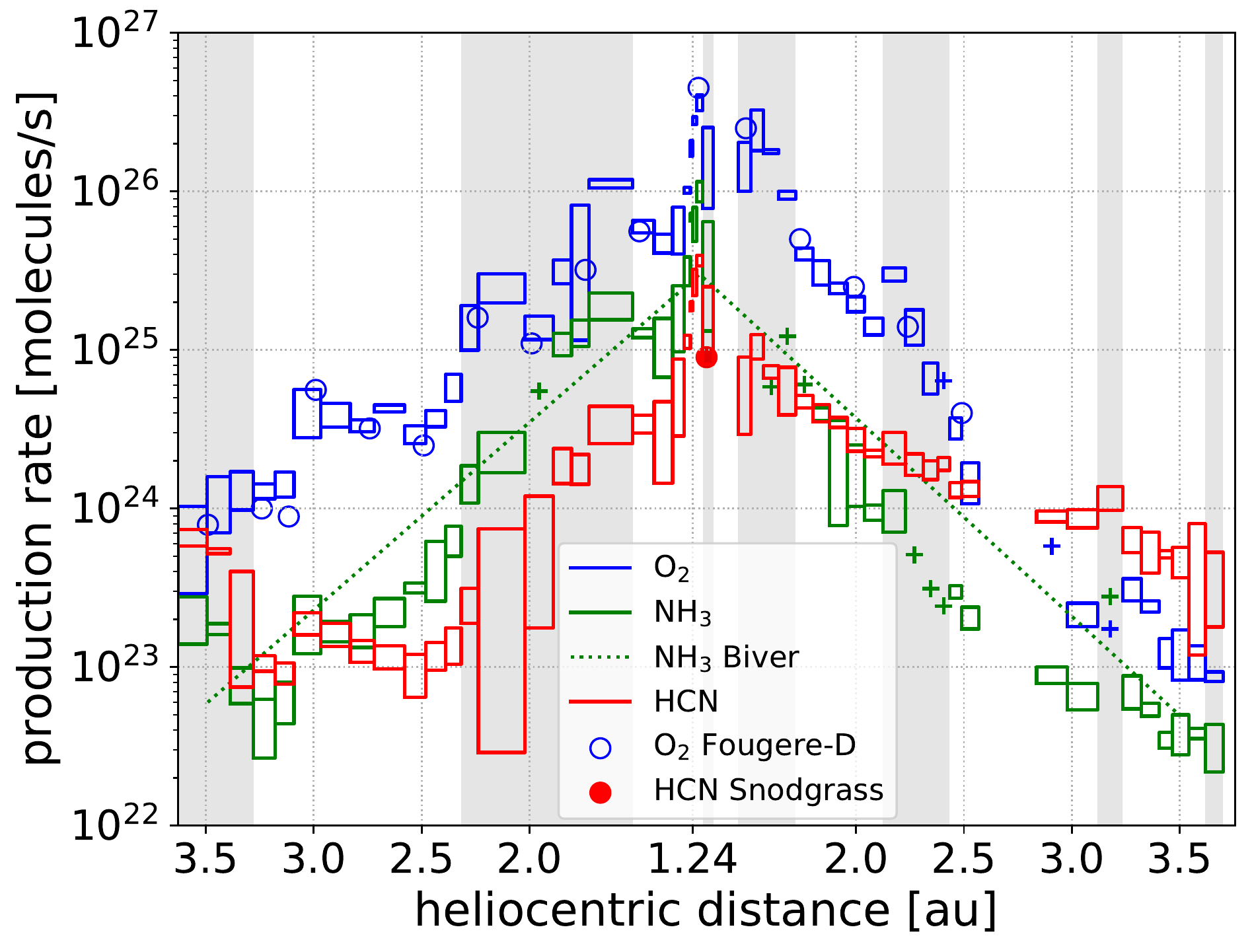}
\caption{Production rates $Q_{s,j}$, the same data as in Fig.~\ref{fig:gastrees}. Top panel: species $s=$ H$_2$O, CO.
Bottom left panel: species $s=$ CO$_2$, C$_2$H$_6$, CH$_3$OH.
Bottom right panel: species $s=$ O$_2$, NH$_3$, HCN. Comparison to the data by {\protect\cite{Biver2019}}, {\protect\cite{Snodgrass2017}}, {\protect\cite{Shinnaka2017}, {\protect\cite{Hansen2016}}, {\protect\cite{Fougere2016}}, {\protect\cite{Fougere2016a}} (D -- DFMS data, V -- VIRTIS data), and {\protect\cite{Fink2016}}.}
}
\label{fig:bivertrees}
\end{figure*}
To classify the evolution of the production rates with heliocentric distance $r_\mathrm{h}$, we fit the production to a power law, $Q_{s,j} \sim r_\mathrm{h}^\alpha$.
Due to the sharp peak production and the sensitivity for chosen fit periods, various authors obtain differing power law exponents.
\cite{Hansen2016} introduced a fit with a discontinuous jump at perihelion, while \cite{Biver2019} employed different fit parameters that changed at $r_\mathrm{h}=1.52\,\mathrm{au}$ outbound.
For three time intervals $I_\mathrm{a}$, $I_\mathrm{b}$, and $I_\mathrm{c}$ in Table~\ref{tab:intervals} our power law fits are performed separately, see Table~\ref{tab:powerfit}.
All chosen intervals correspond to spacecraft terminator-orbits to ensure comparable illumination conditions across the data points.
The vast majority of fits yields negative exponents $\alpha$ corresponding to the increasing production with increasing solar irradiation. 
CO and HCN are two exceptions within the inbound interval $I_\mathrm{a}$.
This inversion of the production rate with respect to the received radiation is clearly visible in Fig.~\ref{fig:gastrees}.
CO$_2$, H$_2$S, O$_2$, and C$_2$H$_6$ remain nearly constant at that time.
\cite{Fougere2016}, \cite{Combi2020} and the 3$\sigma$-points for CO of \cite{Biver2019} in their Fig.~17 observe a similar inversion for CO, see Fig.~\ref{fig:bivertrees}.
For comet C/1995 O1 Hale-Bopp, \cite{Biver2002} report increasing CO and stagnating HCN productions in the same inbound heliocentric distance range from $3$~au to $2$~au.
The explanation by \cite{Enzian1999} focused on interacting sublimations of two different gas species.
For CO$_2$, CO, and HCN the inversion is even more pronounced on the northern hemisphere, see Table~\ref{tab:localpower}.
In contrast, O$_2$, H$_2$S, and C$_2$H$_6$ are less affected by this trend of differences between both hemispheres.

For H$_2$O, O$_2$, H$_2$S, CH$_4$, and NH$_3$ in $I_\mathrm{c}$ our power law exponents closely agree with \cite{Gasc2017}, considered for northern and southern hemisphere separately.
Looking at the longer term trend after perihelion in $I_\mathrm{c}$ we discern two groups of volatiles, the CO$_2$ group and the H$_2$O group in Table~\ref{tab:powerfit}.
The CO$_2$ group is characterized by a slowly decaying production.
In interval $I_\mathrm{c}$, CO$_2$, CO, H$_2$S, C$_2$H$_6$, CH$_4$, HCN, C$_2$H$_5$OH, OCS, and CS$_2$ show similar exponents $-3\leq\alpha$.
This is in contrast to the behaviour in interval $I_\mathrm{b}$ where the gases of the CO$_2$ group show a steeper decrease.
With the exception of C$_2$H$_6$ ($\alpha=-1.6$), $\alpha$ ranges between $-3$ and $-2$ in interval $I_\mathrm{c}$.
The H$_2$O group of gases, namely H$_2$O, O$_2$, CH$_3$OH, H$_2$CO, and NH$_3$, features exponents $\alpha\leq -4.5$ in $I_\mathrm{c}$ and points to a non-linear correlation of the observed production and the received radiation.
H$_2$O, O$_2$, and H$_2$CO are the only gases having a steeper decay for interval $I_\mathrm{c}$ than for interval $I_\mathrm{b}$.
The sublimation of H$_2$O rapidly diminishes towards the end of the spacecraft mission and O$_2$ might be partially trapped in water ice.

\begin{figure*}
\includegraphics[width=0.45\textwidth]{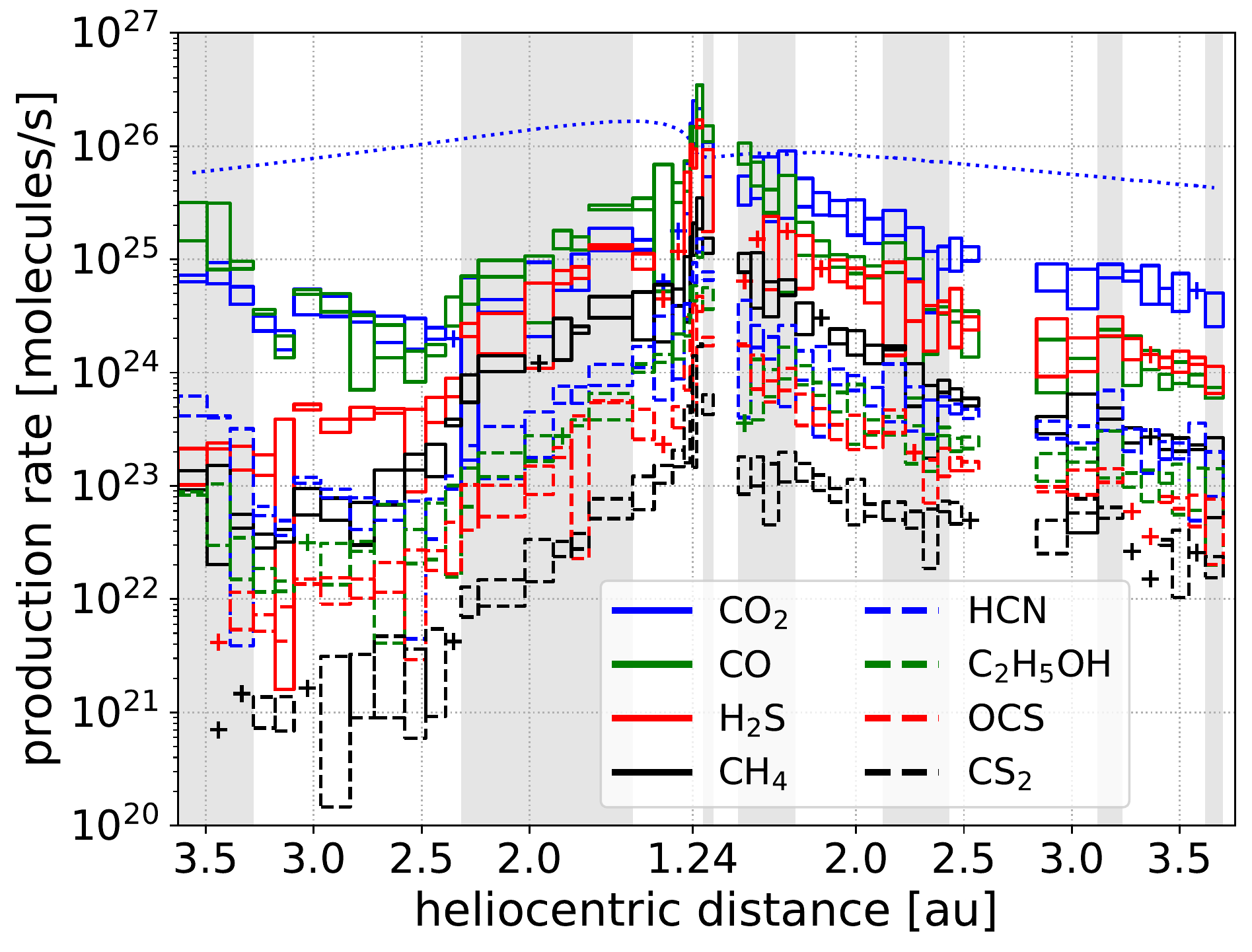}
\includegraphics[width=0.45\textwidth]{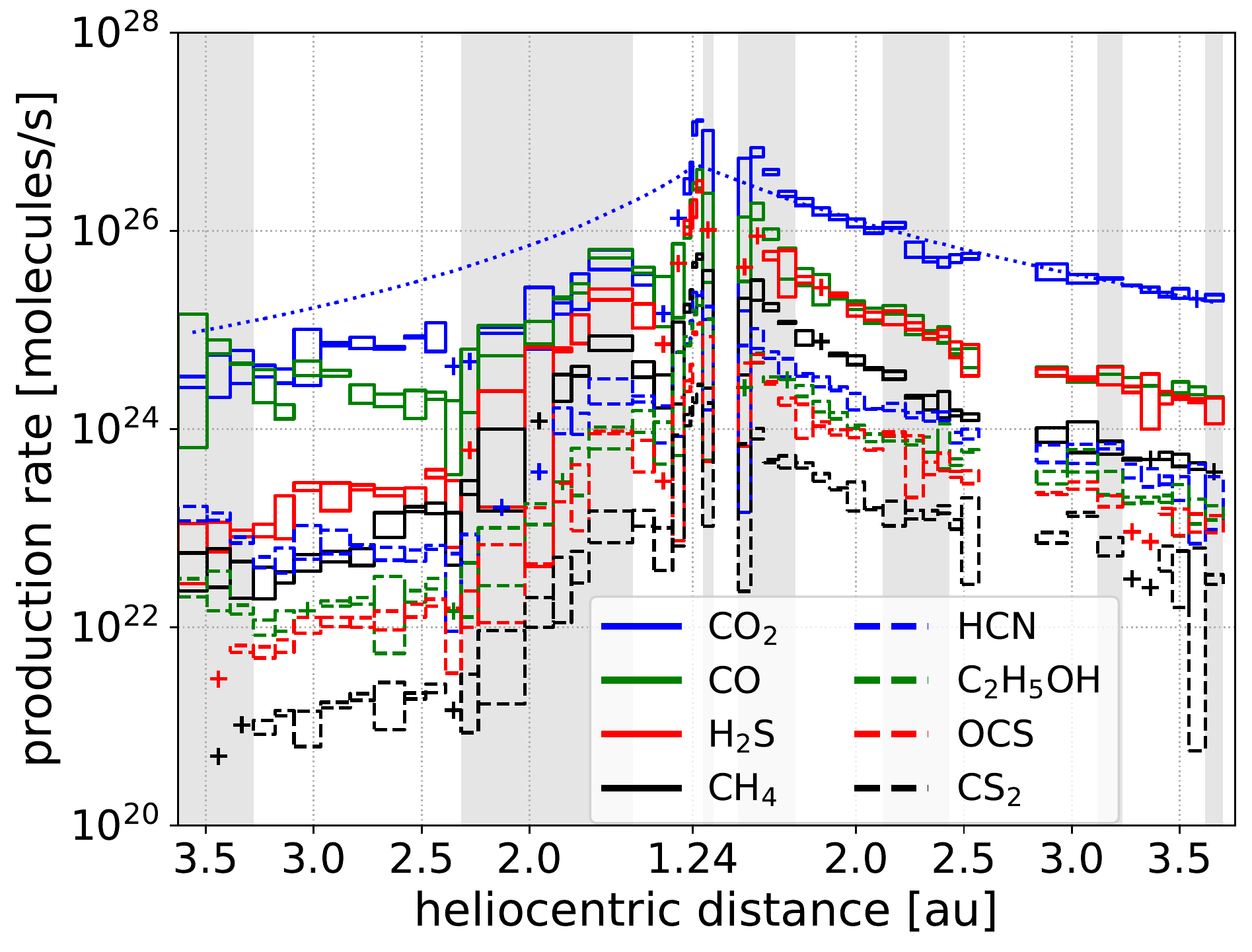}
\caption{Production rates $Q_{s,j}$ for the species $s=$ CO$_2$, CO, H$_2$S, CH$_4$, HCN, C$_2$H$_5$OH, OCS, and CS$_2$. {Left panel: northern hemisphere. Right panel: southern hemisphere.} The idealized radiation driven production in Eq.~\eqref{equ:illu} is shown with dotted lines.}
\label{fig:localpower}
\end{figure*}
It is instructive to compare the observed production to an idealized production model, where the gas production is directly proportional to the instantaneous solar irradiation 
\begin{equation}\label{equ:illu}
Q^{\mathrm{rad}}_{s,j}(A) = \frac{c_s}{r_{\mathrm{h}}^2}
\sum_{f\subset A} |f| \overline{\nu_f \cdot \nu_{\mathrm{sun}}}.
\end{equation}
$A$ denotes an area on the cometary surface consisting of a group of surface elements $f\subset A$, $\nu_f$ is the unit outward vector on $f$, $\nu_{\mathrm{sun}}$ is the instantaneous solar direction vector.
Diurnal averaging is indicated by the bar.
In Fig.~\ref{fig:localpower} we contrast the radiation driven idealized productions with the observed ones.
We have chosen the constant $c_s$ such that the idealized production $Q^{\mathrm{rad}}_{s,25}(A_\mathrm{67P})$ on the complete surface $A_\mathrm{67P}$ at the interval $I_{25}$ (peak production) accounts for half of the observed peak production.
On the northern hemisphere $A_\mathrm{N}$, the decreasing heliocentric distance is partly compensated by the north-south transition of the subsolar latitude in combination with the complex shape of the nucleus.
This effect leads to a smaller slope for $Q^{\mathrm{rad}}_{s,j}(A_\mathrm{N})$ compared to a purely heliocentric distance $r_{\mathrm{h}}^{-2}$ law, see the left panel of Fig.~\ref{fig:localpower}.
{Especially for CO$_2$, CO and HCN} the increasing solar irradiation is not in line with the decreasing productions around $3\,\mathrm{au}$ (inbound), neither for the northern hemisphere nor for the entire surface.
The decreased production could be linked to a different surface morphology and composition, where the comet sheds its accumulated dust from the last perihelion, see \cite{Schulz2015}.
The effect of vertical energy exchange in the surface layer is described in \cite{Gundlach2020} and \cite{Fulle2019}.
For the southern hemisphere $A_\mathrm{S}$ the idealized production $Q^{\mathrm{rad}}_{s,j}(A_\mathrm{S})$ changes faster than $r_h^{-2}$ due to the peculiar shape of the nucleus.
Looking at the gases on the right panel of Fig.~\ref{fig:localpower}, namely CO$_2$, CO, H$_2$S, CH$_4$, HCN, C$_2$H$_5$OH, OCS, and CS$_2$ (the CO$_2$ group except C$_2$H$_6$), for $r_{\mathrm{h}} > 2\,\mathrm{au}$ we find a high correlation between their observed outbound production and the incoming radiation ($\sim Q^{\mathrm{rad}}_{s,j}(A_\mathrm{S})$).
The productions of the species in the H$_2$O group decay much faster  after perihelion in the outbound orbital arc and differ from any idealized production $Q^{\mathrm{rad}}_{s,j}(A_\mathrm{S})$ or a $r_h^{-2}$ relation.
For the inbound intervals the gases of the CO$_2$ group show reduced gas productions (including inversions as described above) compared to $Q^{\mathrm{rad}}_{s,j}(A_\mathrm{S})$.
A separate analysis of the observed production of the small and big lobe does not reveal any differences with respect to the exponents of the production curves, see Table~\ref{tab:localpower} for the exemplary interval $I_\mathrm{a}$.

\section{Gas production for known mission segments}\label{sec:comp}

\begin{figure*}
\includegraphics[width=0.45\textwidth]{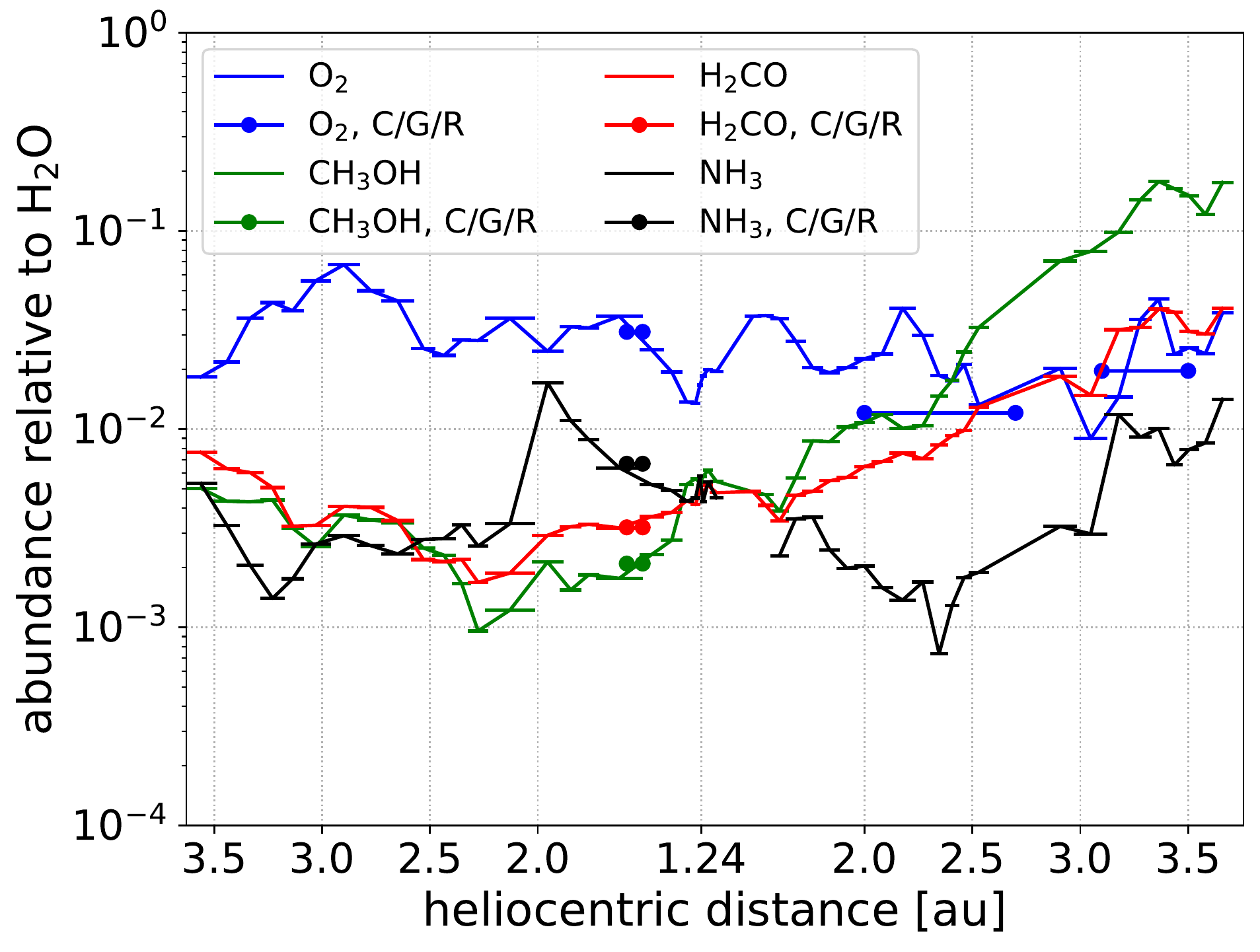}
\includegraphics[width=0.45\textwidth]{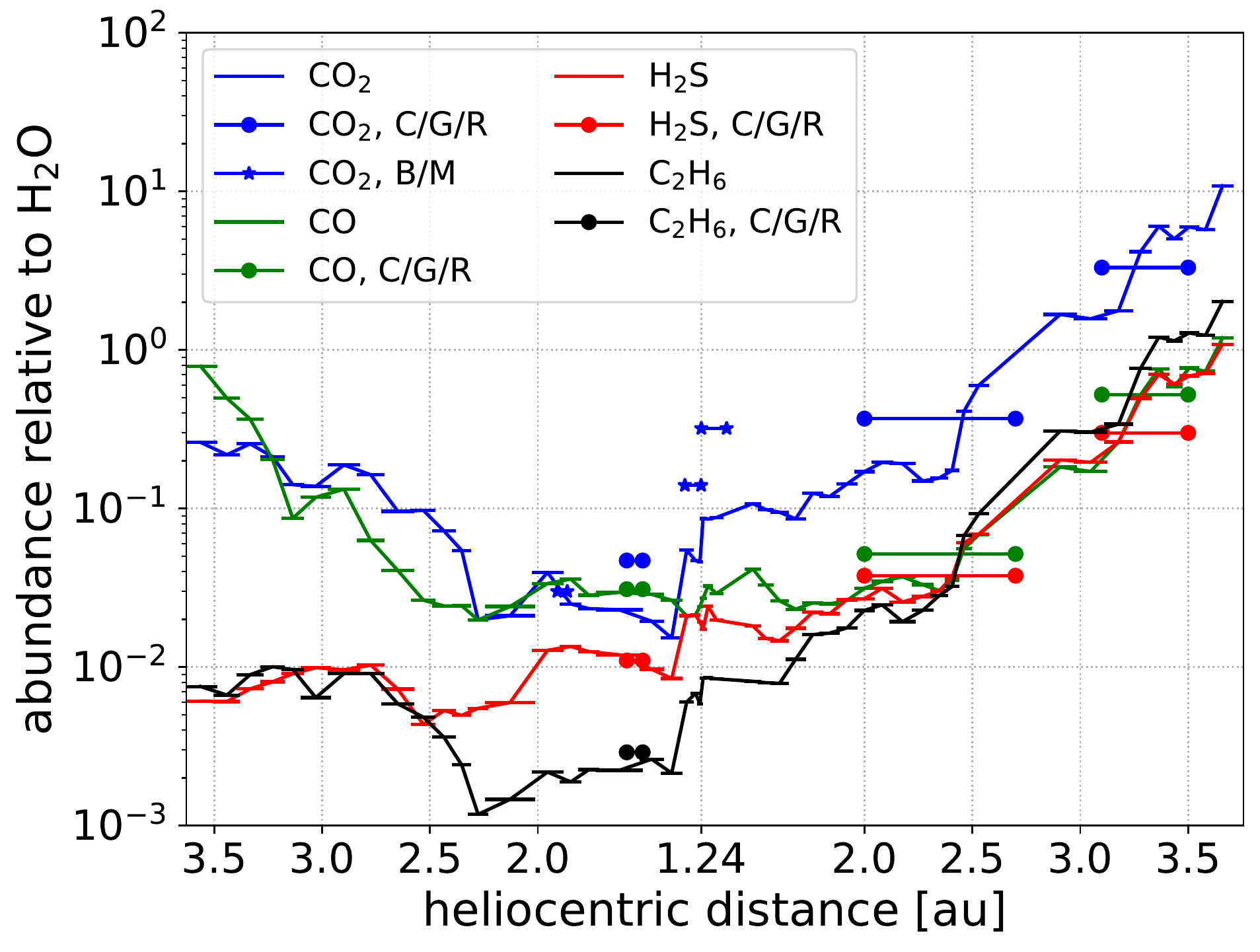}\\
\includegraphics[width=0.6\textwidth]{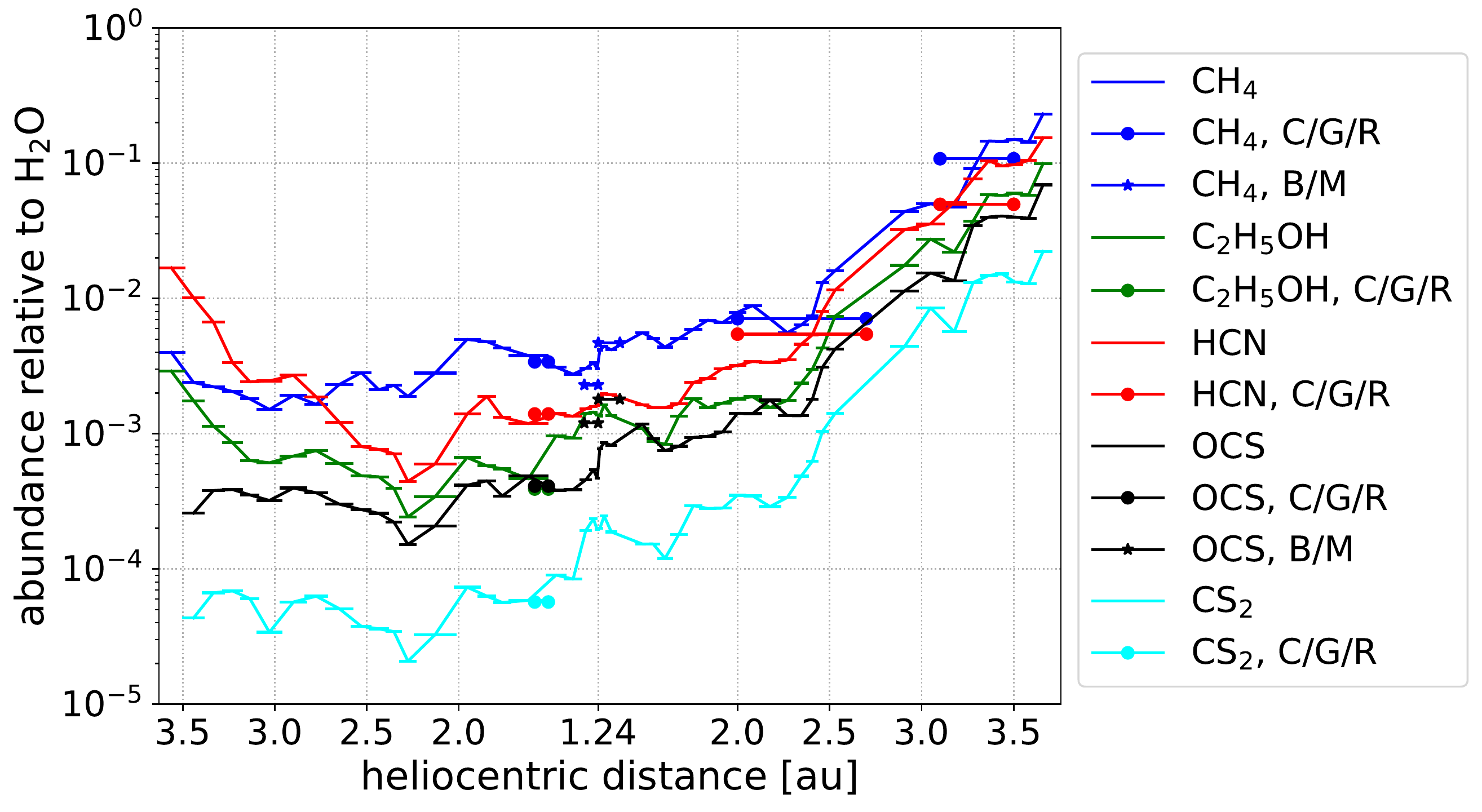}
\caption{Abundances relative to H$_2$O. The horizontal lines reflect the values within each interval $I_j$. The centers of the lines are connected to guide the eye. Data from {\protect\cite{Calmonte2016}}, {\protect\cite{Gasc2017}}, and {\protect\cite{Rubin2019}} are denoted by the shortcut C/G/R. Data from {\protect\cite{Bockelee-morvan2016}} and {\protect\cite{Migliorini2016}} are denoted by the shortcut B/M.}
\label{fig:abundance}
\end{figure*}

In the literature there is a large number of reports concerning the gas production of comet 67P/Churyumov-Gerasimenko.
Our results compare well to known production ranges especially for major gas species.

Before we discuss the minor species we look at the peak production for water 3 weeks after perihelion.
Small deviations from \cite{Lauter2019} reflect differences in the choice of time intervals and in the contribution of not-seen surface areas.
The water production in the Tables~\ref{tab:prod} and \ref{tab:peak} (time-integrated production $P_{\mathrm{H}_2\mathrm{O}}= [ 4.0\pm 0.6 ] \times 10^{9}\,\mathrm{kg}$, peak production $\max_{j} Q_{\mathrm{H}_2\mathrm{O},j} = [ 1.85\pm 0.03 ] \times 10^{28}\,\mathrm{s}^{-1}$) is similar to the values reported by other authors as discussed by \cite{Lauter2019}.
\cite{Lauter2019} already compared $P_{\mathrm{H}_2\mathrm{O}}$ and $\max_{j} Q_{\mathrm{H}_2\mathrm{O},j}$ to the results of \cite{Hansen2016} (based on COPS data), \cite{Marshall2017} (based on MIRO data), and \cite{Shinnaka2017} (based on hydrogen Lyman $\alpha$ data).
Our peak water production is bracketed by the value $\approx 0.5\times 10^{28}\,\mathrm{s}^{-1}$ from \cite{Fougere2016} for VIRTIS data, $0.8\times 10^{28}\,\mathrm{s}^{-1}$ of \cite{Biver2019} for MIRO data, $2.8\times 10^{28}\,\mathrm{s}^{-1}$ in \cite{Combi2020} for DFMS data, and $\approx 3.5\times 10^{28}\,\mathrm{s}^{-1}$ in \cite{Fougere2016} for DFMS data.
\cite{Fougere2016} discuss possible reasons for this data range.
\cite{Bertaux2014} find the peak production 15 days after perihelion for the apparitions 1996, 2002, and 2009 with peak water productions of $1.3 \times 10^{28}\,\mathrm{s}^{-1}$, $1.7 \times 10^{28}\,\mathrm{s}^{-1}$, and $5.65 \times 10^{27}\,\mathrm{s}^{-1}$, respectively.
The time-integrated water production $4.9\times 10^{9}\,\mathrm{kg}$ by \cite{Combi2020} falls close to our uncertainty range.
The lower water production based on MIRO data is also notable for the integrated production between $2.42\times 10^9\,\mathrm{kg}$ and $3.3\times 10^9\,\mathrm{kg}$ in \cite{Biver2019}.

Fig.~\ref{fig:bivertrees} shows the temporal evolution of production rates for 8 selected species (H$_2$O, CO$_2$, CO, O$_2$, C$_2$H$_6$, CH$_3$OH, NH$_3$, and HCN) complemented by the results of other authors.
This comparison shows the agreement with the published fits of \cite{Biver2019}, \cite{Hansen2016}, \cite{Combi2020}, and  \cite{Fougere2016,Fougere2016a} in the range of uncertainties discussed by \cite{Hansen2016}.
Fig.~\ref{fig:bivertrees} also includes the water production given by \cite{Snodgrass2017} for the end of July 2015 ($5.1\times 10^{27}\,\mathrm{s}^{-1}$), which is within our error bounds whereas for the week after perihelion their value $3.2\times 10^{27}\,\mathrm{s}^{-1}$ is lower by a factor of 4 compared to our estimation.
At later times, (September, October and November 2015) their values are again within our error bounds.
Our CO$_2$ curves resembles the one by \cite{Fougere2016}, also derived from DFMS data, with the exception of the peak production.
Our peak CO$_2$ production given in Table~\ref{tab:peak} is bracketed by the peak value of VIRTIS data ($\approx 1\times 10^{27}\,\mathrm{s}^{-1}$) and the DFMS data ($\approx 6\times 10^{27}\,\mathrm{s}^{-1}$) from \cite{Fougere2016}.
The CO$_2$ production between 3\,au and 2.4\,au inbound derived from VIRTIS data by \cite{Fougere2016a} fluctuates considerably (see Fig.~\ref{fig:bivertrees}), with the higher values agreeing with our results.
The $Q_{\mathrm{CO}_2,j}$ value $\approx 1.2\times 10^{25}\,\mathrm{s}^{-1}$ between February and April 2015 in \cite{Fink2016} (based on VIRTIS data) underestimates our values at that time.
CO shows a close relation to the DFMS data in \cite{Fougere2016}, exceptions are their higher values for peak production, at 2.0\,au, and at 2.5\,au.
The CO fit of \cite{Biver2019} strongly underestimates our DFMS derived production.
However, 7 out of the 13 $3\sigma$-limit values reported by \cite{Biver2019} are close to our lower bound estimate.
The O$_2$ production derived from the DFMS data by \cite{Fougere2016} agrees with our values.
Their peak production value for CO$_2$ exceeds our value, while  their peak value for O$_2$ is closer to our result.
Out of perihelion the values are in good agreement.
The productions in \cite{Snodgrass2017}, $9.9\times 10^{25}\,\mathrm{s}^{-1}$ for C$_2$H$_6$ (July 2015), $9\times 10^{24}\,\mathrm{s}^{-1}$ for HCN (September 2015), and $2\times 10^{26}\,\mathrm{s}^{-1}$ for CH$_3$OH (September 2015) correspond closely to our results.
Our productions of NH$_3$, inbound and outbound, and of CH$_3$OH outbound, agree  with \cite{Biver2019}.

Besides the confirmation of absolute gas productions also the linked relative abundances agree with the results of other authors for various volatiles.
The relative abundances with respect to H$_2$O are shown in Fig.~\ref{fig:abundance}.
The relative abundances are in agreement (less than 30\% deviation) with the DFMS analysis by \cite{Rubin2019} for CO, H$_2$S, O$_2$, CH$_3$OH, H$_2$CO, NH$_3$, HCN, OCS in May 2015, by \cite{Gasc2017} for O$_2$ in two time intervals, January -- March 2016 and June -- July 2016, and with the Fig.~12 in \cite{Fougere2016} for CO$_2$, CO, and O$_2$ in the time interval August 2014 -- February 2016.
The CO$_2$ abundance in April 2015 of $\approx 0.03$ by \cite{Migliorini2016} is also reproduced by our analysis.
The relative abundances of CO$_2$, CH$_4$, and OCS during July/August 2015 and August/September 2016 differ from the numbers reported by \cite{Bockelee-morvan2016}, whereas the qualitative evolution of $Q_{s}/Q_{\mathrm{H}_2\mathrm{O}}$ for $s=\mathrm{CO}, \mathrm{CH}_3\mathrm{OH}, \mathrm{NH}_3$ by \cite{Biver2019} is mirrored by our results.
Between 200 days before and after perihelion CO shows an almost constant ratio with respect to water, CH$_3$OH strongly increases, and NH$_3$ decreases.
\cite{Bockelee-morvan2017} review further abundances relative to water on other comets. 
In particular O$_2$ is strongly linked to the water production with maximum deviations of $Q_{\mathrm{O}_2}/Q_{\mathrm{H}_2\mathrm{O}}$ in the range $0.009$ -- $0.07$ during the entire mission.
Toward the end of mission all other relative abundances increase and reflect the steep decrease of the water production.

\section{Conclusions}\label{sec:conclusions}

The sublimation of cometary ices fuels the coma of comet 67P/Churyumov-Gerasimenko with a variety of volatiles.
Based on COPS/DFMS data from the {\it Rosetta} spacecraft mission and an inverse gas model the temporal evolution of the gas production for 14 species has been reconstructed and investigated.
This includes the detection of outliers with a $2\sigma$ criterion for the $l^2$-error functional.
Our results compare well to previous publications using data from the same and other instruments (COPS, DFMS, MIRO, and VIRTIS).
This concerns the time-integrated production for the complete mission, peak production rates for major and minor gas species, especially for water, and relative abundances relative to water.

Increasing solar radiation toward perihelion leads to a long term trend of increased gas production for all species with a peak production in the time interval between day 17 and day 27 after perihelion.
Because the temporal evolution for gas productions on the two lobes does not show significant differences we do not see an indication for different ice compositions on both lobes.
A similar finding has been reported by \cite{Schroeder2019}, who found the same deuterium-to-hydrogen ratio in H$_2$O above the two lobes.

During the outbound times between $190\,\mathrm{d}$ ($2.4\,\mathrm{au}$) and $380\,\mathrm{d}$ ($3.6\,\mathrm{au}$) (interval $I_\mathrm{c}$)
the gas production for the southern hemisphere shows a strong correlation with the solar irradiation for the species CO$_2$, CO, H$_2$S, CH$_4$, HCN, C$_2$H$_5$OH, OCS, and CS$_2$.
This points to an almost linear coupling between solar irradiation and sublimation rate, similar to the assumptions behind Model~A in \cite{Keller2015a}.
The power law exponents obtained by \cite{Gasc2017} (-2.18, -1.83, and -2.76) for the southern production of CO$_2$, CO, and HCN confirm this finding.
Complemented by C$_2$H$_6$ this group of species coincides with the CO$_2$ group which is given by the property $-3\leq\alpha$ for the exponent of the global production in the outbound interval $I_\mathrm{c}$.

We observe three phenomena with a more complex relation between solar irradiation and gas production.
The first observation concerns the significant production decrease for the gases CO and HCN during the interval $I_\mathrm{a}$ between $-290\,\mathrm{d}$ ($3.1\,\mathrm{au}$) and $-180\,\mathrm{d}$ ($2.3\,\mathrm{au}$) before perihelion and increasing irradiation at the same time.
During the same time period the production for the species CO$_2$, H$_2$S, O$_2$, and C$_2$H$_6$ does not increase.
This finding extends the observed decoupling of gas production from solar irradiation by \cite{Biver2002} for CO on comet C/1995 O1 Hale-Bopp.
The second point refers to the H$_2$O group of gases, defined by the exponent $\alpha\leq -4.5$ of the global production in interval $I_\mathrm{c}$.
The slopes of the H$_2$O group are much steeper compared to the irradiation decrease after perihelion.
A third result is the analysis of the gas production from the northern hemisphere.
There, we did not find a strong correlation between solar radiation and gas production.
This points to differences in the sublimation properties and thus ice decomposition of the northern and the southern hemisphere.

All three observations point to complex relations between solar radiation and gas production.
Physical processes explaining this observation need to overcome present assumptions like diurnally averaged irradiation, a sublimation function depending on the instantaneous irradiation (without diurnal or seasonal delay), and fixed surface properties.
\cite{Gundlach2020}, \cite{Skorov2020} and \cite{Fulle2019} describe non-linear interactions in the soil column which might explain such effects.

Our data processing with the automatic detection of outliers based on a $2\sigma$ criterion within each interval $I_j$ (in Sect.~\ref{sec:data}) excludes data from COPS/DFMS which deviates strongly from the diurnally averaged gas production of the coma model.
The outlier analysis could be used in future work to identify short lasting event, for instance outbursts on the cometary surface.

\section*{Acknowledgements}

The work was supported by the North-German Supercomputing Alliance
(HLRN).
{\it Rosetta} is an ESA mission with contributions from its member states
and NASA.
We acknowledge herewith the work of the whole ESA {\it Rosetta} team.
ROSINA would not have produced such outstanding results without the work of the many engineers, technicians, and scientists involved in the mission, in the {\it Rosetta} spacecraft team and in the ROSINA instrument team over the past 20 years, whose contributions are gratefully acknowledged.
Work on ROSINA at the University of Bern was funded by the State of
Bern, the Swiss National Science Foundation, and by the European Space
Agency PRODEX program.

\section*{Data availability}

{The time series of the global production for all 14 volatiles shown in Fig.~\ref{fig:gastrees} is available in CSV format as ancillary files to} \url{https://arxiv.org/abs/2006.01750}.

\label{lastpage}

\end{document}